\documentstyle[12pt]{article}
\def\TL{\hfil$\displaystyle{##}$}
\def\TR{$\displaystyle{{}##}$\hfil}

\def\TT{\hbox{##}}
\def\seqalign#1#2{\vcenter{\openup1\jot
  \halign{\strut #1\cr #2 \cr}}}

\def\fixit#1{}

\def\mop#1{\mathop{\rm #1}\nolimits}

\def\vol{\mop{vol}}

\def\Tr{\mop{Tr}}

\def\overleftrightarrow#1{\vbox{\ialign{##\crcr
     $\leftrightarrow$\crcr\noalign{\kern-0pt\nointerlineskip}
     $\hfil\displaystyle{#1}\hfil$\crcr}}}

\def\lsim{\mathrel{\mathstrut\smash{\ooalign{\raise2.5pt\hbox{$<$}\cr\lower2.5pt\hbox{$\sim$}}}}}
\def\gsim{\mathrel{\mathstrut\smash{\ooalign{\raise2.5pt\hbox{$>$}\cr\lower2.5pt\hbox{$\sim$}}}}}

\def\sqr#1#2{{\vcenter{\vbox{\hrule height.#2pt
         \hbox{\vrule width.#2pt height#1pt \kern#1pt
            \vrule width.#2pt}
         \hrule height.#2pt}}}}
\def\square{\mathop{\mathchoice\sqr56\sqr56\sqr{3.75}4\sqr34\,}\nolimits}

\def\href#1#2{#2}  

\def\lbldef#1#2{\expandafter\gdef\csname #1\endcsname {#2}}
\def\eqn#1#2{\lbldef{#1}{(\ref{#1})}%
\begin{equation} #2 \label{#1} \end{equation}}
\def\eqalign#1{\vcenter{\openup1\jot
    \halign{\strut\span\TL & \span\TR\cr #1 \cr
   }}}
\def\eno#1{(\ref{#1})}

\begin{document}
\baselineskip=15.5pt
\pagestyle{plain}
\setcounter{page}{1}
\begin{titlepage}

\begin{flushright}
CALT-68-2345 \\
CITUSC/01-030 \\
PUPT-2005 \\
hep-th/0108239
\end{flushright}
\vfil

\begin{center}
{\huge Some interesting violations}
\vskip0.5cm
{\huge of the Breitenlohner-Freedman bound}
\end{center}

\vfil
\begin{center}
{\large Steven S. Gubser$^{1,2}$ and Indrajit Mitra$^{2}$}
\end{center}

$$\seqalign{\span\TL & \span\TT}{
^1 & Lauritsen Laboratory of Physics, 452-48 Caltech, Pasadena, CA  91125  \cr
^2 & Joseph Henry Laboratories, Princeton University, Princeton, NJ 08544
}$$
\vfil

\begin{center}
{\large Abstract}
\end{center}

\noindent 
 We demonstrate that $AdS_5 \times T^{pq}$ is unstable, in the sense
of Breitenlohner and Freedman, for unequal $p$ and $q$.  This settles,
negatively, the long-standing question of whether the $T^{pq}$
manifolds for unequal $p$ and $q$ might correspond to
non-supersymmetric fixed points of the renormalization group.  We also
show that the $AdS_3 \times S^7$ vacuum of Sugimoto's $USp(32)$ open
string theory is unstable.  This explains, at a heuristic level, the
apparent absence of a heterotic string dual.

\vfil
\begin{flushleft}
August 2001
\end{flushleft}
\end{titlepage}
\newpage

\section{Introduction}
\label{Introduction}

The Breitenlohner-Freedman (BF) bound \cite{BF} (see also
\cite{mt,bfTwo,townsend}) says that tachyons in anti-de Sitter space
(AdS) cause an instability only if their mass-squared falls below a
certain negative threshold, of order the reciprocal of the AdS radius
squared.  In the AdS/CFT correspondence \cite{juanAdS,gkPol,witHolOne}
(for a review see \cite{MAGOO}), AdS vacua with some field(s)
violating the BF bound need not have a well-defined field theory dual.
Indeed, if one attempts to compute the two-point function of such a
field, the result is highly cutoff-dependent.  This is like having a
lattice theory without a well-defined continuum limit.  By extension,
solutions to string theory or supergravity which are asymptotic to AdS
vacua violating the bound may also be expected to have some pathology
on the field theory side.

Supersymmetry guarantees that the BF bound will be satisfied, but it
is well-known that the converse is not true.  For instance, there is
an infinite family of compactifications of M-theory, $AdS_4 \times
M^{pqr}$, which are stable but non-supersymmetric
\cite{ppTwo}.\fixit{Try to figure out which of these to cite where.}
$M^{pqr}$ is a homogeneous Einstein 7-manifold describable as a coset
space $(SU(3) \times SU(2) \times U(1)) /(SU(2) \times U(1) \times
U(1))$. The integers $p$, $q$, and $r$ describe the embedding of
$SU(2) \times U(1) \times U(1)$ in $SU(3) \times SU(2) \times
U(1)$. In most cases the symmetry group of these spaces is $SU(3)
\times SU(2) \times U(1)$.\footnote{The two exceptions are $M^{101} =
S^5 \times S^2$ and $M^{011} = CP^2 \times S^3$.}  For a certain range
of $p$, $q$, and $r$, there is a BF-violating tachyon, and for the
complementary range there is not.

A question many times raised, but never (as far as we know) seriously
investigated, is whether a similar phenomenon occurs for $AdS_5 \times
T^{pq}$ compactifications.  It is well-understood that the $AdS_5
\times T^{11}$ case is stable, since it is the supersymmetric
near-horizon limit of D3-branes on a conifold \cite{kwOne}.  Also,
$AdS_5 \times T^{kk}$ must be stable (at least classically) since it
is a smooth, supersymmetry-breaking ${\bf Z}_k$ quotient of $T^{11}$, as we
shall review further below.  On the other hand, $T^{01}$ is a direct
product space, $S^2 \times S^3$, and such product geometries are
always unstable toward inflating one factor while deflating the other,
provided the total number of compact dimensions is less than nine
\cite{dnp,dfghm}.  The question then is whether $T^{pq}$ is stable for
some range of $p/q$ close to $1$, as occurs in the $M^{pqr}$ case.  In
fact it is not: we shall demonstrate that all $T^{pq}$ for $p \neq q$
are unstable by constructing the unstable mode explicitly.  To a large
extent this dashes the hope that renormalization group flows from the
simplest ${\bf Z}_2$ orbifold of four-dimensional ${\cal N}=4$ gauge
theory might include infinitely many infrared fixed
points.\footnote{The $M^{pqr}$ manifolds are topologically distinct
from one another for different values of ${p \over q}$, so there is no
question of whether one could flow from one to another.}
 (There is still the
possibility that stable solutions exist with topology
$S^2 \times S^3$ and with three-form field strengths and non-trivial
dilaton; however these seem much more difficult to find).

A second question which we address is the stability of the $AdS_3
\times S^7$ vacuum of Sugimoto's $USp(32)$ open string theory
\cite{Sugimoto}.  The reason to be interested in this vacuum is that
it is the near-horizon limit of many coincident D1-branes in this
theory.  For the usual $SO(32)$ open string, the D1-brane turns out to
be a non-perturbative construction of the $SO(32)$ heterotic string
\cite{hs,pw}.  There is no perturbative $USp(32)$ heterotic string in
ten dimensions: the $USp(32)$ current algebra is too big to admit
unitary representations with $c \leq 16$.  Correspondingly, it is
perhaps satisfying that we find fields which violate the BF bound in
the $AdS_3 \times S^7$ vacuum of Sugimoto's theory.  One may perhaps
draw the general conclusion that string theories which are
non-supersymmetric in their perturbative construction can suffer
non-perturbative instabilities which prevent them from participating
in weak-strong coupling dualities.

The organization of the paper is as follows.  After some introductory
remarks on Freund-Rubin compactifications in section~\ref{Freund}, we
present the stability analysis for the $T^{pq}$ backgrounds in
section~\ref{Stability}. In section~\ref{Operator} we discuss candidate
field theory operators of dimension $2$ which are dual to the modes of the
supergravity analysis which just saturate the stability bound for $T^{11}$.
After this, we proceed to the $AdS_3 \times S^7$
background of $USp(32)$ open string theory in section~\ref{Sugimoto}.

\section{Freund-Rubin Backgrounds}
\label{Freund}

In this section, for completeness and to fix notation, we review the
Freund-Rubin background.  We start out by considering classical
$D=p+q$ dimensional gravity theory coupled to a $q$-form field
strength.  The action is given by:
 \eqn{Action}{
 S = \int{d^{p}x d^{q}y \sqrt{-g} \left(R - {1 \over {2 q!}} F_{q}^2 \right)} \,,
  }
which leads to the equations of motion: 
 \eqn{MetricEOM}{
R_{MN} = {1 \over {2 (q-1)!}} F_{M P_2 \cdots P_q} F^{\;\; P_2 \cdots P_q}_N - {{(q-1)} \over {2 (D-2) q!}} g_{MN}F_{q}^2 \,,
 }
 \eqn{FormEOM}{
 d * F_{q} =0 \,.
 }
This theory supports a Freund-Rubin solution with the product metric:
 \eqn{ProdSpace}{
 ds^2 = ds_{AdS_p}^2 + ds_{M_q}^2 \,,
 } 
describing a product of $p$-dimensional anti-de Sitter space with an Einstein manifold:
 \eqn{BackgroundAdS}{
 R_{\mu \nu} = -{{(p-1)} \over L^2}g_{\mu \nu} \,,
 }
 \eqn{BackgroundM}{
 R_{\alpha \beta} = {{(q-1)} \over R^2}g_{\alpha \beta} \,, 
 }
and a background field strength on the compact space:
 \eqn{BackgroundF}{
 F_{q} = e \vol(M_q) \,.
 }
We use $M, N, \ldots$ for indices on the full $D$-dimensional
spacetime, while $\mu, \nu, \ldots$ are indices on $AdS$ and $\alpha, \beta, \ldots$ are indices on $M_q$. The equations of motion
(\ref{MetricEOM}), (\ref{FormEOM}) relate the length scales
$L$ and $R$ and the constant $e$:
 \eqn{Valueofc}{
 e^2 = {{2(D-2)(q-1)} \over {(p-1) R^2}} \,,
}
 \eqn{RadiusRatio}{
 L = {{p-1} \over {q-1}} R \,.
 }

\section{Stability Analysis for Compactifications on $T^{pq}$}
\label{Stability}

Let us now examine the issue of stability of Type IIB supergravity
compactified on the manifold commonly known as $T^{pq}$. This is a
five-dimensional Einstein manifold which is a coset of $SU(2) \times
SU(2)$ by a $U(1)$ whose generator can be written as $p\Sigma_3 + q
\tilde\Sigma_3$, where $\Sigma_3$ and $\tilde\Sigma_3$ are generators
of the two $SU(2)$'s.  The integers $p$ and $q$ describe the winding
numbers of the $U(1)$ fiber over the two spheres.  The most general
metric on $T^{pq}$ consistent with $SU(2) \times SU(2) \times U(1)$
isometry is:
 \eqn{MetricT}{
 ds^2 = a^2 (dy_1^2 + \sin^2 y_1 dy_2^2) + b^2 (dy_3^2 + \sin^2 y_3 dy_4^2) + c^2 (dy_5 - p \cos y_1 dy_2 - q \cos y_3 dy_4)^2 \,,
}
 where $a$, $b$, and $c$ are constants, $y_1$ and $y_3$ range from $0$
to $\pi$, $y_2$ and $y_4$ range from $0$ to $2 \pi$, and $y_5$ ranges
from $0$ to $4 \pi$.  Conditions on $a$, $b$, and $c$ for the metric
\MetricT\ to be Einstein were discussed in \cite{romans}, and we will
recap some of the relevant points as they will apply to our subsequent
analysis.  We will assume throughout that $p$ and $q$ are relatively
prime, and then in the last paragraph of this section address the case
where they are not.

Let us choose the following basis of $1$-forms:
  \eqn{Basis}{\eqalign{
   E^1 &= a dy_1 \,,\quad
   E^2 = a \sin y_1 dy_2 \,,\quad
   E^3 = b dy_3 \,,\quad      
   E^4 = b \sin y_3 dy_4 \,, \cr
   E^5 &= c(dy_5 - p \cos y_1 dy_2 - q \cos y_3 dy_4) \,.
}}
The spin-coefficients in this basis are:
\eqn{Spin}{\eqalign{
 \omega_{12} &= - {1 \over a} \cot y_1 E^2 - {{pc} \over {2a^2}} E^5  \,,\quad 
 \omega_{13} = \omega_{14} = 0 \,,\quad
 \omega_{15} = -{{pc} \over {2a^2}} E^2 \,,\quad
 \omega_{25} = {{pc} \over {2a^2}} E^1 \,, \cr
 \omega_{34} &= - {1 \over b} \cot y_3 E^4 - {{qc} \over {2b^2}} E^5 \,,\quad
 \omega_{23} = \omega_{24} = 0 \,,\quad
 \omega_{35} = - {{qc} \over {2b^2}} E^4 \,,\quad
 \omega_{45} = {{qc} \over {2b^2}} E^3 \,.
 }}
The curvature components are calculated using the relation:
\eqn{Curv}{
R^{\mu}{}_{\nu} = d \omega^{\mu}{}_{\nu} + \omega^{\mu}{}_{\alpha} \wedge \omega^{\alpha}{}_{\nu} \,.
}
Only a few of them need to be actually computed. The remaining ones can be
found using the symmetry of the metric and the symmetric and anti-symmetric
properties of the Riemann tensor. However, for completeness we list all of
the components of the curvature $2$-form:
\eqn{RComp}{\eqalign{
 R^{1}{}_{2} &= \left( {1 \over a^2} - {{3p^2c^2} \over {4a^4}} \right) E^1 E^2 - {{pqc^2} \over {2 a^2 b^2}}E^3 E^4 \,,\quad 
 R^{1}{}_{3} = -{{pqc^2} \over {4 a^2 b^2}} E^2 E^4 \,,\quad
 R^{1}{}_{4} = {{pqc^2} \over {4 a^2 b^2}} E^2 E^3 \,, \cr
 R^{3}{}_{4} &= \left( {1 \over b^2} - {{3q^2c^2} \over {4b^4}} \right) E^3 E^4 - {{pqc^2} \over {2 a^2 b^2}}E^1 E^2 \,,\quad
 R^{2}{}_{4} = -{{pqc^2} \over {4 a^2 b^2}} E^1 E^3 \,,\quad
 R^{2}{}_{3} = {{pqc^2} \over {4 a^2 b^2}} E^1 E^4 \,, \cr
 R^{1}{}_{5} &= {{p^2 c^2} \over {4a^4}}E^1 E^5 \,,\quad
 R^{2}{}_{5} = {{p^2 c^2} \over {4a^4}}E^2 E^5  \,,\quad 
 R^{3}{}_{5} = {{q^2 c^2} \over {4b^4}} E^3 E^5 \,,\quad
 R^{4}{}_{5} = {{q^2 c^2} \over {4b^4}} E^4 E^5 \,.
}}
Finally, we have to demand that the metric above is Einstein. In the
orthonormal basis that we chose above, this condition is simply that
$R^{i}{}_{j} = \Lambda \delta^{i}{}_{j}$, where $\Lambda$ is the constant
of proportionality between the Ricci tensor and the metric. This yields
three equations relating the constants $\Lambda$, $a$, $b$ and $c$:

\eqn{Einstein}{
\Lambda = {{2 a^2 - p^2 c^2} \over {2 a^4}} = {{2 b^2 - q^2 c^2} \over {2 b^4}}  = {{(a^4 q^2 + b^4 p^2) c^2} \over {2 a^4 b^4}} \,.
}
For convenience, let us work in units where the radius of one of the spheres is set equal to unity, i.e. $ a \equiv 1$. In these units the other constants $b$ and $c$ are:
\eqn{Valuebc}{
b^2 = {1 \over {3 \Lambda -1}} \quad \quad c^2 = {{2(1- \Lambda)} \over p^2} \,.
} 
It is also helpful to express the ratio ${q \over p}$ in terms of $\Lambda$:
\eqn{Valuer}{
 \left( {q \over p} \right)^2 = {{2 \Lambda -1} \over {(1 - \Lambda)(3 \Lambda - 1)^2}} \,.
}
Looking at the last expression we see that $\Lambda$ varies between ${1
\over 2}$ and $1$. So, given any manifold $T^{pq}$ we first evaluate the
ratio ${q \over p}$ and then using (\ref{Valuer}) compute $\Lambda$. For
instance, the space $T^{11}$ has $\Lambda = {2 \over 3}$. All questions
about stability can be answered in terms of values of
$\Lambda$.

In \cite{dfghm} it was shown that for an arbitrary Einstein manifold, the
masses of the scalar modes resulting from a mixing of 3 scalars: the trace
of the metric on $AdS_5$, the trace on $M_5$ and another scalar which
arises from the fluctuations of the five form field strength  never violate
the stability bound - they can at most saturate it. The masses of the
coupled scalar modes are:
 \eqn{Safemode}{
  m^2 L^2 = \lambda + 16 \pm 8 \sqrt{\lambda + 4} \,,
 }
 where $\square_y Y \equiv - \lambda Y/R^2$ and $Y$ is a scalar
harmonic on $T^{pq}$.  By $\square_y$ we will always mean
$\nabla^\alpha \nabla_\alpha$.  Minimizing with respect to $\lambda$ we find
that the least massive mode corresponds to $\lambda = 12$.  Moreover,
this mode just saturates the stability bound $m^2 L^2 \geq -4$. The
isometry group of $T^{pq}$ is $SU(2) \times SU(2) \times U_R(1)$ so
the eigenvalues of the scalar Laplacian on $T^{pq}$ are expressed in
terms of the eigenvalues $j_1, j_2, r$ corresponding to the two
$SU(2)$'s and the $U(1)_R$ \cite{Jatkar,SSG,Ferrara}:
 \eqn{ScalarLap}{
  {\lambda \over R^2} = {r^2 \over c^2} + {1 \over a^2} \left[ j_1 (j_1 +1) - (pr)^2 \right] + {1 \over b^2} \left[ j_2 (j_2 +1) - (qr)^2 \right] \,,
 }
with $j_1 \geq pr$ and $j_2 \geq qr$. Let us examine this for the special
case of $T^{11}$. Here we have from (\ref{Valuer}) and (\ref{Valuebc})
$\Lambda = {2 \over 3}$, $a=b=1$, and $c^2 = {2 \over 3}$. Since $\Lambda =
{4 \over R^2}$ from (\ref{BackgroundM}) and from (\ref{RadiusRatio}) $R=L$,
the expression for the eigenvalue of the scalar harmonic on $T^{11}$
simplifies to:
\eqn{ScalarOneOne}{
 \lambda = 6 \left[ j_1 (j_1 + 1) + j_2 (j_2 + 1) - {r^2 \over 2} \right] \,.
}
The value of $\lambda =12$ is thus satisfied for $(j_1,j_2,r)= (1,0,0)$ and $(0,1,0)$.   

None of these coupled scalar modes can violate the stability bound for any
$T^{pq}$. But, it was shown in \cite{dfghm} that the only mode which could
potentially violate the stability bound is the traceless graviton mode.
Without proof again, we write down the equation of motion for this
fluctuation (for a derivation, the reader is referred to
\cite{dfghm}):
\eqn{curvature}{
 [(\square_x + \square_y) \delta^{a}_{c} \delta^{b}_{d} - 2
   R_{c}{}^{ab}{}_{d} ] \left( \phi (x) Y_{ab} (y) \right) = 0 \,,
}
 where as usual, $\square_y = \nabla^\alpha \nabla_\alpha$ and
$\square_x = \nabla^\mu \nabla_\mu$.  This equation may be rewritten
in terms of the Ricci tensor and the Lichnerowicz operator $\Delta_L$. The action of the latter on symmetric tensors $Y_{ab}$ is defined as:
\eqn{DefLich}{
  \Delta_L Y_{ab} \equiv \square_y Y_{ab} - 2 R_{a}{}^{cd}{}_{b}
    Y_{cd} - 2 R_{(a}{}^{c}Y_{b)c} \,.
}
 Here and below, $(\ldots)$ indicates symmetrization: $(ab) = (ab+ba)/2$.
Using the definition \DefLich, the fluctuation equation for the symmetric traceless graviton modes (\ref{curvature}) is simply:
  \begin{eqnarray}
   [ \square_x + \Delta_L + {2 (q-1) \over R^2} ] 
    \left( \phi Y_{ab} \right) &=& 0 \,. \label{Lphieqn}
  \end{eqnarray}
 Since we're dealing with $AdS_5 \times M_5$, $p=q=5$ and from
(\ref{RadiusRatio}) we have $R=L$.  The Breitenlohner-Freedman bound
is $m^2 L^2 \geq -4$. Assembling all these facts together we can
translate the BF bound from a bound on the mass to one on the
eigenvalue of the Lichnerowicz operator acting on symmetric tensors
$Y_{ab}$:
  \eqn{LichOp}{
   \Delta_L Y_{ab} = \lambda Y_{ab} \,.
  }
 For stability we must have:
\eqn{BFTwo}{
\lambda L^2 \leq -4 \,.
}
It is somewhat painful to diagonalize the Lichnerowicz operator
directly.  Fortunately, a trick employed in \cite{ppTwo} works here as
well.\footnote{We thank C.~Pope for a
communication which brought this paper to our attention.} The key is to use the identity:
  \eqn{Identity}{
   \int dV \, Y^{ab} \Delta_L Y_{ab} = 
    \int dV \, [-4Y^{ab}R_{acdb}Y^{cd} - 4 \Lambda Y^{ab}Y_{ab} -
     3 \nabla^{(a} Y^{bc)} \nabla_{(a} Y_{bc)}] \,,
  }
 which can easily be demonstrated by writing out explicitly $Y^{ab}
\Delta_L Y_{ab}$ and $\nabla^{(a} Y^{bc)} \nabla_{(a} Y_{bc)}$,
simplifying, and integrating by parts. Let us diagonalize the Riemann
tensor by solving the eigenvalue equation:
 \eqn{RiemannDiag}{
R_{a}{}^{bc}{}_{d} Y_{bc} = \kappa Y_{ad} \,.
}
This would involve diagonalizing a $15 \times 15$ matrix. One of the
eigenvectors would be pure trace, so that we'll be left with $14$ traceless
eigenvectors. Using (\ref{BFTwo}) and (\ref{Identity}) our stability bound
now reads: The geometry would be stable if every $\kappa$
satisfies:
\eqn{BFThree}{
\kappa \geq  {1 \over L^2} - \Lambda \,.
}
{}From (\ref{BackgroundM}) we have $\Lambda = {4 \over R^2}$ and since $R=L$, the above bound can be expressed solely in terms of $\Lambda$ as:
\eqn{FinalBF}{
\kappa_{min} \geq - {3 \over 4} \Lambda \,,
}
 where $\kappa_{min}$ is the least of the $14$ eigenvalues.  Thus we
have reduced the problem of solving the complicated equation
(\ref{curvature}) into a simple one of diagonalizing the Riemann
tensor. On account of the simple metric and the symmetries involved,
there is very little mixing of the modes and the problem is
sufficiently simple to be solved by hand. The eigenvectors and
eigenvalues are shown in table~\ref{TableOne}.
  \begin{table}[ht]
  \begin{center}
  \begin{tabular}{|c|c|c|}
  \hline
   Eigenvectors & Eigenvalues & Eigenvalues in units $a=1$ \\[8pt]
  \hline
   $X_{ab}^{12}$, $X_{ab}^{11} - X_{ab}^{22}$ & $\displaystyle{{1 \over a^2} - {{3 p^2 c^2} \over {4 a^4}}}$ & ${1 \over 2}(3 \Lambda -1)$ \\[8pt]
   $X_{ab}^{34}$, $X_{ab}^{33} - X_{ab}^{44}$ & $\displaystyle{{1 \over b^2} - {{3 q^2 c^2} \over {4 b^4}}}$ & ${1 \over 2}$  \\[8pt]
   $X_{ab}^{13}+X_{ab}^{24}$, $X_{ab}^{14} + X_{ab}^{23}$ & $\displaystyle{{{3pqc^2} \over {4a^2 b^2}}}$ & ${3 \over 2} \sqrt{(1-\Lambda)(2 \Lambda -1)}$ \\[8pt]
   $X_{ab}^{13}-X_{ab}^{24}$, $X_{ab}^{14} - X_{ab}^{23}$ & $\displaystyle{-{{3pqc^2} \over {4a^2 b^2}}}$ & $ -{3 \over 2} \sqrt{(1-\Lambda)(2 \Lambda -1)}$ \\[8pt]
   $X_{ab}^{15}$, $X_{ab}^{25}$ & $\displaystyle{{{p^2 c^2} \over {4a^4}}}$ & ${1 \over 2} (1 - \Lambda)$ \\[8pt]
   $X_{ab}^{35}$, $X_{ab}^{45}$ & $\displaystyle{{{q^2 c^2} \over {4b^4}}}$ & $(\Lambda - {1 \over 2})$ \\[8pt]
   $\alpha_{+} (X_{ab}^{11} + X_{ab}^{22}) + \beta_{+} (X_{ab}^{33} + X_{ab}^{44}) + \gamma_{+} X_{ab}^{55}$ & & ${1 \over 4}(- \Lambda + \sqrt{49 \Lambda^2 - 60 \Lambda + 20})$ \\[8pt]
   $\alpha_{-} (X_{ab}^{11} + X_{ab}^{22}) + \beta_{-} (X_{ab}^{33} + X_{ab}^{44}) + \gamma_{-} X_{ab}^{55}$ & & ${1 \over 4}(- \Lambda - \sqrt{49 \Lambda^2 - 60 \Lambda + 20})$ \\[8pt]
  \hline
  \end{tabular} 
  \caption{Eigenvectors and eigenvalues of the Riemann tensor, as
defined in \RiemannDiag.}\label{TableOne}
  \end{center}
  \end{table}
 In this table, $\alpha_{\pm}$, $\beta_{\pm}$ and $\gamma_{\pm}$ are
constants given by
 \eqn{EigVec}{\eqalign{
\alpha_{\pm} &= 4 (1 - 2 \Lambda)(7 \Lambda - 4 \mp \sqrt{49 \Lambda^2 - 60 \Lambda + 20})  \,, \cr
\beta_{\pm}  &= 2 (1 - 2 \Lambda)(- 7 \Lambda + 2  \pm \sqrt{49 \Lambda^2 - 60 \Lambda + 20}) \,, \cr
\gamma_{\pm} &= (2 - \Lambda \pm \sqrt{49 \Lambda^2 - 60 \Lambda + 20})(- 7 \Lambda + 2  \pm \sqrt{49 \Lambda^2 - 60 \Lambda + 20}) \,,
}}
 and we have also defined
  \eqn{XDef}{
   X_{ab}^{cd} = \delta_{(a}^c \delta_{b)}^d \,.
  }
 We have not written down the eigenvalues for the last two modes in
the table in terms of $a$, $b$, $c$, $p$ and $q$ because the
expressions are quite lengthy. Remembering that ${1 \over 2} \leq
\Lambda \leq 1$ we find that there is only one mode (the last one in
the table) which can violate the bound (\ref{FinalBF}). For this
potentially dangerous mode we find that only $\Lambda = {2 \over 3}$
 (which corresponds to the manifold $T^{11}$) saturates the bound, while all other values of $\Lambda$ lead to
masses which violate the bound. This tells
us that the only stable compactification on $T^{pq}$ manifolds with
$p$ and $q$ relatively prime turns
out also to be the only one which preserves supersymmetry. (Note that
the modes $X_{ab}^{13} - X_{ab}^{24}$ and $X_{ab}^{14} - X_{ab}^{23}$ saturate the bound
for $T^{11}$ while for all other $T^{pq}$ have eigenvalues which are
above the bound. This might lead us to suspect that for $T^{11}$ these
modes have masses $m^2 L^2 = -4$. But on careful examination we find
that they do not satisfy the Killing tensor equation $\nabla_{(a}
Y_{bc)} = 0$. Therefore, according to (\ref{Identity}) they have
masses $m^2 L^2 > -4$). 

A word now about what the unstable mode (or in the case of $T^{11}$ the
marginally stable mode) looks like. For $T^{11}$, we can use (\ref{EigVec})
to evaluate the constants $\alpha_{-}= - {8 \over 3}$, $\beta_{-}= {8 \over
3}$, and $\gamma_{-} = 0$ so that the eigenvector which just saturates the
bound is simply diag$(-1,-1,1,1,0)$. Geometrically this is a fluctuation in
which one $S^2$ expands while the other shrinks with the length of the
$U(1)$ fiber unchanged. For generic $T^{pq}$ however, such a simple picture
is not obtained. As an example, let us consider the unstable mode of
$T^{12}$. For this manifold, using (\ref{Valuer}) we find $\Lambda \approx
0.9331$, and using (\ref{EigVec}) we have $\alpha_{-} \approx -17.73$,
$\beta_{-} \approx 12.33$, and $\gamma_{-} \approx 10.80$ so this
fluctuation makes one $S^2$ shrink and the other expand accompanied by an
elongation of the fiber.

To summarize, for $T^{11}$ we have found a
total of seven modes which saturate the stability bound.  Six of them
come from the coupled scalar modes, and the remaining one is a
traceless graviton mode. All of these modes have masses $m^2 L^2 =
-4$.  Using the relation between the mass and the dimension of the
corresponding operators in the dual field theory, $\Delta = {1 \over
2}[(p-1) \pm \sqrt{(p-1)^2 + 4 m^2 L^2}]$ (here $p$ is the dimension
of $AdS$), we find that the operators have scaling dimension $2$. In
the next section we shall examine in some detail the issue of
identifying these operators in the dual field theory according to the
AdS/CFT correspondence.

In the above analysis we have only shown that if the inequality
(\ref{FinalBF}) is satisfied, then stability is guaranteed. Let us now
prove that if the inequality is violated, then we necessarily have
instability in the traceless graviton sector. Looking at
(\ref{Identity}) we find that we have to demonstrate that the putative
unstable mode is also a Killing tensor obeying $\nabla_{(a} Y_{bc)}
=0$. To prove this, we make the following observations. First, if we
restrict all three indices $a$, $b$, and $c$ to lie in the four
manifold which are the two $S^2$'s of $T^{pq}$ (i.e. these indices are
allowed to run from $1$ to $4$), then a constant, diagonal tensor of
the form:
 \eqn{IntMode}{
Y_{ab} = \rm{diag} (\alpha, \alpha, \beta, \beta, \gamma) \,,
}
with $2 \alpha + 2 \beta + \gamma =0$ is covariantly constant. Second, we notice that the spin-coefficients of the metric written down in (\ref{Spin}) can be split up in the following way:
\eqn{SpinSplit}{
 \omega^{5}_{ab} = \omega^{4}_{ab} + c_{ab} E^{5} \qquad
 \omega^{5}_{a5} = c_{ab} E^{b} \,,
}
where $\omega^{5}$ refers to the full spin connection, $\omega^{4}$ refers
to the part on the two $S^2$'s, and $c_{ab}$ is antisymmetric in $a$ and
$b$. Using this fact and the form of the constant diagonal tensor, one can
show that indeed this eigentensor satisfies the Killing condition. So we
have demonstrated that the only stable $AdS_5 \times T^{pq}$
compactification with $p$ and $q$ relatively prime is on $T^{11}$.

To extend the discussion to the case of $p$ and $q$ not relatively
prime, a topological point should be made first: if ${\rm gcd}(p,q) =
k \neq 1$, then $T^{pq}$ is topologically $S^2 \times S^3/{\bf Z_k}$,
where the ${\bf Z}_k$ acts freely on the Hopf fiber of $S^3$.  The
$k=1$ case of this statement follows based on arguments given in
\cite{candelas}; the $k>1$ statement follows as a corollary when one
notes that modding out by the $U(1)$ generated by $p\Sigma_3 +
q\tilde\Sigma_3$ can be accomplished by first modding out by the
$U(1)$ generated by $(p\Sigma_3 + q\tilde\Sigma_3)/k$ and then
dividing by ${\bf Z}_k$.  Moreover, $T^{pq}$ is metrically a quotient
of $T^{p/k,q/k}$ by ${\bf Z}_k$ acting on the $U(1)$ fiber.  Thus one
can flow without encountering a topological obstruction from any
$T^{pq}$ to any other precisely when ${\rm gcd}(p,q)$ remains
unchanged.  In each class of manifolds $T^{pq}$ with ${\rm
gcd}(p,q)=k$ fixed, there is precisely one which is classically
stable, namely $T^{kk}$.  Only for $k=1$ is any supersymmetry
preserved.  The perturbation analysis on $T^{pq}$ could be carried out
by considering ${\bf Z}_k$ invariant functions on $T^{p/k,q/k}$.  The
orbifolding by $k$ also has a well-defined meaning on the gauge theory
side, resulting for $T^{kk}$ in a theory with gauge group $SU(N)^{2k}$
and some complicated matter.  Details of counting and the operator map
could be pursued for the case of general $k$, but in the next section
we will do so only for $k=1$.

\section{The Operator Map for $T^{11}$}
\label{Operator}

 In the previous section we saw that the Freund-Rubin compactification of
Type IIB on $AdS_{5} \times T^{11}$ have three modes which have masses
which just saturate the stability bound. Recall that two of these modes
came from the coupled scalars and the remaining one is a traceless graviton
mode. According to the AdS/CFT correspondence, these modes should
correspond to operators whose dimension is protected and would therefore be
either chiral primaries or conserved currents. So let us try to find the
dual operators.  
 
 The compactification of Type IIB SUGRA on $T^{11}$ has $SU(2,2|1)$
symmetry. Let us try to put these fluctuations into $SU(2,2|1)$
supermultiplets. A unitary highest weight representation of $SU(2,2|1)$ can
be
decomposed into a direct sum of unitary highest weight representations
of the bosonic subalgebra $SU(2,2) \times U_R(1)$ whose maximal
compact subalgebra is $U(1) \times SU(2) \times SU(2) \times U_R(1)$
\cite{FGPW, FlatoF}. The first $U(1)$ is the energy and the last one is the
$R$-charge. So the highest weight representations of $SU(2,2|1)$ are
labelled by four quantum numbers $D(E_0,s_1,s_2;r)$. In addition to
these, there are of course, the quantum numbers associated with the
symmetry of the isometry group of $T^{11}$ which is $SU(2) \times
SU(2) \times U_R(1)$. Note that this $U_R(1)$ is the same $U_R(1)$ as
the one associated to the R-charge. We'll call the additional quantum
numbers due to these last two $SU(2)$'s as $(j_1,j_2)$. So our first
task is to find out all the six quantum numbers of the fluctuation
mode in question $(E_0, s_1, s_2; r, j_1, j_2)$.

 From the $5d$ AdS point of view all of the modes in question,
including $Y_{ab}$, are scalars. So, the spin quantum numbers
$s_1=s_2=0$. To compute the AdS energy, we use the relation between
mass and energy for a scalar in $5d$ AdS space:
 \eqn{Energy}{
 E_0 = 2 \pm \sqrt{4 + m^2 L^2} \,.
} 
We found above that these modes have masses $m^2 L^2 = -4$, so $E_0 = 2$.
Finally, to obtain the value of $r$, we note that for a representation of
$SU(2,2|1)$ to be unitary, there are inequalities among the four quantum
numbers $E_0$, $s_1$, $s_2$, and $r$. The relevant one for our purposes
here is $E_0 \geq 2 s_2 + {3 \over 2} r + 2$ which fixes
$r=0$.
We observe now that this set of quantum numbers $(2,0,0,0)$ satisfies $3$ multiplet shortening conditions \cite{FGPW} (which is what is expected for a field saturating the unitarity bound):
\eqn{ShortCond}{\eqalign{
 E_0 - 2 s_1 + {3 \over 2} r -2 = 0 \quad \quad
 E_0 - 2 s_2 + {3 \over 2} r -2 = 0 \quad \quad
 s_2 = 0 \,.
}}
So we get the following multiplet with only $4$ fields present:

\begin{center}
\begin{tabular}{|c|c|c|c|}
\hline
 $E_{0}/R$ & $r=-1$ & $r=0$ & $r=1$ \\[8pt]
\hline
 $2$           &                 & $(0,0)$                      &                 \\[8pt]
 ${3 \over 2}$ & $({1 \over 2},0)$ &                            & $(0,{1 \over 2})$ \\[8pt]
 $3$           &                 & $({1 \over 2}$, ${1 \over 2})$ & 		      \\[8pt]
\hline
\end{tabular}
\end{center}

\noindent
where the quantities in the table refer to the quantum numbers $(s_1, s_2)$. For completeness, we note that the masses of these fields can be calculated using the relations: 
\eqn{MultiMass}{\eqalign{
  \left( {1 \over 2}, 0 \right) \left( 0, {1 \over 2} \right) \quad \quad m &=  |{E_0 -2}| = {1 \over 2} \,, \cr
  \left( {1 \over 2}, {1 \over 2} \right) \quad \quad m^2 &= (E_0 - 1)(E_0 - 3) = 0 \,.
}}

Let us now turn to the field theory realization of this. The conformal
field theory dual to the supergravity theory has two doublets of
chiral superfields $A_i,B_j \; (i,j = 1,2)$ transforming in the $(N,
\bar N)$ and $(\bar N, N)$ representations of $SU(N) \times
SU(N)$. These fields both have $R$-charge $1$. The global symmetry
group $SU(2) \times SU(2)$ quantum numbers for these fields are $({1
\over 2}, 0)$ and $(0, {1 \over 2})$, respectively. According to the
AdS/CFT correspondence, each supergravity field with quantum numbers
$(E_0, s_1, s_2;r)$ is mapped to a conformal field with scaling
dimension $\Delta = E_0$, Lorentz quantum numbers of an $SL(2,C)$
representation $(s_1,s_2)$, and an $R$-symmetry charge $r$. Since we
had multiplet shortening, we know that the dimension of the
corresponding superfield would be protected, and furthermore we also
determined its dimension to be $2$.  There are natural field theory
candidates with the desired properties to be dual to the scalars we
have found.\footnote{We thank M.~Strassler for useful communications
which helped us identify the field theory operators.}  Namely,
consider the real superfields:\footnote{In \eno{CurrentA}, $V_1$ and
$V_2$ are the real superfields that include the $SU(N) \times SU(N)$
gauge fields.  There is a notational subtlety: $A^*_i$ transforms as a
doublet of $SU(2)$, and we have omitted the $\epsilon_{ij}$ which
would usually be inserted to make the group action come out right.}
  \eqn{CurrentA}{\eqalign{
   J_A &= \Tr A_{(i} e^{V_2} A^{*}_{j)} e^{V_1} \qquad 
    J_B = \Tr B_{(i} e^{V_1} B^{*}_{j)} e^{V_2}  \cr
   J_{\rm baryon} &= \Tr A_{[1} e^{V_2} A^*_{2]} e^{V_1} - 
    \Tr B_{[1} e^{V_1} B^*_{2]} e^{V_2} \,.
  }}
 The vector component of each of these is a conserved current, by
Noether's theorem: $J_A$ is associated with the global $SU(2)$
rotating $A_1$ and $A_2$; $J_B$ is associated with the other global
$SU(2)$; and $J_{\rm baryon}$ is associated with the unbroken
$U(1)_{\rm baryon}$.  The scalar component of each of these
superfields, call them ${\cal O}_A$, ${\cal O}_B$, and ${\cal O}_{\rm
baryon}$, have protected dimension~$2$ and R-charge $0$.  The operator
${\cal O}_{\rm baryon}$ was discussed in \cite{kwTwo} in the context
of resolving the conifold.  Of the supergravity modes with masses $m^2
L^2 = -4$, we conjecture that the baryon current is the one which is
dual to the traceless graviton mode, while $J_A$ and $J_B$ are dual to
the coupled scalar modes.  The global $SU(2) \times SU(2)$ charges
support this expectation. Moreover, both the traceless graviton fluctuation
that we examined in the previous section and its proposed operator dual
$J_{\rm baryon}$ flip sign on interchanging the $S^2$'s, i.e. they are both
$\bf{Z_2}$ odd.

\section{Sugimoto's $USp(32)$ open string theory}
\label{Sugimoto}

 The next example of an unstable Freund-Rubin compactification that we
shall consider arises in the $USp(32)$ open string theory considered
in \cite{Sugimoto}. As we shall discover, the modes which are unstable
come from a mixing of the trace of the metric on $AdS_3$, the trace on
$S^7$, and another scalar arising from the fluctuations of the form
field. Thus, this instability is of the same type as the one for the
Freund-Rubin compactification of massive Type IIA supergravity on
$AdS_4 \times S^6$ \cite{dfghm}.  The low-energy effective action of
the Sugimoto theory in the string frame is \cite{DudasMourad}:
 \eqn{stringaction}{
S= {1 \over {2 \kappa^2}} \int d^{10} x \sqrt{G} \left[ e^{-2 \phi} \left(R
+ 4 (\partial \phi)^2 \right) - {1 \over 12}F_3^2 - \alpha e^{- \phi}
\right] \,.
}
where $\alpha$ for our purposes is just a constant. In our conventions, $F_3^2 = F_{MNP} F^{MNP}$.  To bring the
action~\stringaction\ into Einstein frame, we rescale the metric as
$g_{MN} = e^{- \phi \over 2} G_{MN}$. The action then becomes:
 \eqn{Einaction}{
S={1 \over {2 \kappa^2}} \int d^{10} x \sqrt{g} \left[R - {1 \over 2}
(\partial \phi)^2 - \alpha e^{{3 \over 2} \phi} - {1 \over 12} e^{\phi}
F_3^2 \right] \,.
}
The scalar equation of motion which follows from this action is:
\eqn{Redscalar}{
\square \phi - {3 \over 2} \alpha e^{{3 \over 2} \phi} - {1 \over 12}
e^{\phi} F_3^2 = 0 \,.
}
For a constant $\phi$ background, this equation can have a solution if we
use the Freund-Rubin ansatz $F_{\mu \nu \rho} = f \epsilon_{\mu \nu\rho}$, 
i.e.{} the three-form is along the $AdS$ part. So we find that $AdS_3 \times S^7$
is indeed a solution with $\phi = 0$.

For convenience, let us dualize the three-form and use a seven-form
instead. The action (which is what we shall be using from now on) is:
 \eqn{useaction}{
S = {1 \over {2 \kappa^2}} \int d^{10} x \sqrt{g} \left[ R - {1 \over 2}
(\partial \phi)^2 - {1 \over {2 \cdot 7!}} e^{- \phi} F_7^2 - 
  \alpha e^{{3 \over 2} \phi} \right] \,.
}
The background geometry has $\phi=0$. The equations of motion are:
\eqn{SugiEinst}{
R_{MN} = {1 \over {2 \cdot 6!}} e^{- \phi} F_{MP_1 \cdots P_5}
  F_{N}{}^{P_1 \cdots P_5} - {3 \over
{8 \cdot 7!}} e^{- \phi}F_7^2 g_{MN} + {1 \over 2} \partial_M \phi \partial_N \phi + {1 \over
8} \alpha e^{{3 \over 2} \phi} g_{MN} \,,
}
\eqn{form}{
d*(e^{- \phi} F_7) = 0 \,,
}
\eqn{scalar}{
\square \phi + {1 \over {2 \cdot 7!}}e^{- \phi} F_7^2 - 
  {3 \over 2} \alpha e^{{3 \over 2} \phi} = 0 \,.
}
 We want to express all the parameters in terms of the $AdS$ radius
$L$.  For the background $F_7 = c \vol_{S^7}$ and $\phi=0$ so (\ref{scalar}) gives
$c^2 = {1 \over 7!} F_7^2 = 3 \alpha$. The Einstein equation yields $\alpha
= {2 \over L^2}= {12 \over {R^2}}$. So, the ratio of the radii is $R^2
= 6 L^2$. Let us now proceed to get the mass spectrum of the scalars.
Tracing over the indices on the sphere in (\ref{SugiEinst}) gives:
 \eqn{SugiRicci}{
R_{\alpha}^{\alpha}= {7 \over {4L^2}}e^{{3 \over 2} \phi} + {7 \over {8 \cdot 7!}}e^{- \phi}F_7^2 + {1 \over 2} \partial^{\alpha} \phi \partial_{\alpha} \phi\,.
}
When expanded to linear order, the two sides of the above equation yield:
\eqn{LinEinstein}{
-{1 \over L^2} \pi - {1 \over 2}(\square_{x} + \square_{y}) \pi = -{{21} \over {8 L^2}} \delta \phi + {7 \over 4} c \square_{y} b - {{21} \over {4L^2}} \pi \,.
}
where $\pi$ denotes the trace of the metric fluctuation on $S^7$, and the
fluctuation of the $7$-form field strength is expressed as $\delta F_7 = d
a_6$ with $a_6 = {*_7} db$. On simplification this finally
gives:
\eqn{Sugione}{
(\square_x + \square_y) \pi - {{17} \over {2L^2}}\pi + 21 \square_y B - {{21} \over {4L^2}} \delta \phi = 0 \,,
}
where we've introduced the notation $b \equiv c L^2 B$.
For the scalar fluctuations we expand (\ref{scalar}) to linear order:
\eqn{Sugitwo}{
(\square_x + \square_y) \delta \phi - {15 \over {2L^2}} \delta \phi - {3 \over L^2} \pi + 6 \square_y B = 0 \,.
}
The form equation (\ref{form}) expanded to linear order yields after a little algebra:
\eqn{Sugithree}{
(\square_x + \square_y)B - {6 \over {7 L^2}} \pi - {1 \over L^2}\delta \phi = 0 \,.
}
Assembling all the three equations, and assuming that $B$, $\pi$, and
$\delta\phi$ are eigenvectors of $\square_y$ with eigenvalue
$-\lambda/R^2$, we obtain the following mass matrix equation:
\eqn{IIAMassMatrix}{\eqalign{
  \square_x \pmatrix{ B \cr \cr \pi \cr \cr \delta\phi} = 
  \pmatrix{ {\lambda \over R^2} & {6 \over {7L^2}} & {1 \over L^2} \cr \cr
   {21 \lambda \over R^2} & {{\lambda \over R^2} + {{17} \over {2 L^2}}} & {{21} \over {4 L^2}} \cr \cr
   {6 \lambda \over R^2} & {3 \over L^2} & {{\lambda \over R^2} + {{15} \over {2 L^2}}}}
  \pmatrix{ B \cr \cr \pi \cr \cr \delta\phi} \,.
 }}
 On diagonalizing the matrix, and using the relation $R^2=6L^2$ to
eliminate $R^2$, we obtain the eigenvalues (mass squared) $m^2 L^2 =
{{\lambda + 24} \over {6}}$, ${{\lambda + 36 + 12 \sqrt{\lambda + 9}} \over {6}}$, and ${{\lambda
+ 36 - 12 \sqrt{\lambda +9}} \over {6}}$.  Only states in the last tower can
be tachyonic.  On $S^7$, the spherical harmonics have eigenvalues
$\lambda = k(k+6)$. The dangerous tower of states when expressed in
terms of $k$ become, $m^2 L^2 = {k(k-6) \over 6}$. Remembering that
the BF bound for this system is $m^2 L^2 \geq -1$, we see that the
modes $k=2$, $3$, and $4$ violate the bound.  The presence of three
unstable modes makes it considerably more difficult to find a stable
compactification where $S^7$ is replaced by some other seven-manifold
$M_7$: there is a fairly wide range of eigenvalues for the laplacian
on $M_7$ which would lead through \IIAMassMatrix\ to an unstable mode.

\section{Conclusions}
\label{Conclusions}

We have investigated two types of anti-de Sitter compactifications
and found them to be unstable in the sense of Breitenlohner and
Freedman.  In the context of AdS/CFT, this instability implies that
neither the AdS background nor solutions asymptotic to it at infinity
have a unitary field theory dual.  To arrive at a solution of
supergravity that does have such a dual, it would be necessary to
change the asymptotics of the spacetime.

The first class of AdS compactifications we studied is $AdS_5 \times
T^{pq}$.  Except for the case $p=q$, we have found
these solutions to be unstable, and the unstable mode which we were
able to construct is a metric fluctuation under which one $S^2$ of
$T^{pq}$ expands, the other shrinks, and the $U(1)$ fiber either
elongates or shrinks by an amount required to keep the total volume of
the manifold constant.  Thus it still 
remains to find a stable, non-supersymmetric anti-de Sitter
compactification of type IIB supergravity which is not locally
isometric to a supersymmetric one.  Infinitely many such
compactifications of eleven-dimensional supergravity to $AdS_4 \times
M_7$ have long been known, as we remarked in the introduction.  This
problem of non-supersymmetric $AdS_5$ vacua takes on a new interest in
light of AdS/CFT, because it corresponds to discovering
four-dimensional, non-supersymmetric, strong-coupling fixed points of
the renormalization group.

The other compactification which we studied is the $AdS_3 \times S^7$
solution to Sugimoto's $USp(32)$ open string theory.  Here the field
theory dual, if it existed, would be the CFT describing many
coincident dual heterotic strings.  A heterotic dual does {\it not}
seem to exist for this non-supersymmetric open string theory, and so
it is sensible that we again find violations of the
Breitenlohner-Freedman bound.  This may lead us to wonder anew to what
extent the duality web of M-theory depends on supersymmetry.

\section*{Acknowledgments}

We would like to thank D.~Freedman, I.~Klebanov, and M.~Strassler for
useful discussions, and C.~Pope for bringing the useful results of
\cite{ppTwo} to our attention.  We also thank D.~Freedman and
C.~Pope for comments on an early draft.  This work was supported in
part by the DOE under grants DE-FG03-92ER40701 and DE-FG02-91ER40671
and through an Outstanding Junior Investigator Award.

\bibliography{unstable}
\bibliographystyle{ssg}

\end{document}